\documentclass[conference,compsoc]{IEEEtran}

\usepackage{graphicx}
\graphicspath{ {images/} }
\usepackage[font=small,skip=0pt]{caption}
\usepackage{tabu}
\usepackage{multirow}
\usepackage{array}
\newcolumntype{L}[1]{>{\raggedright\let\newline\\\arraybackslash\hspace{0pt}}m{#1}}
\newcolumntype{C}[1]{>{\centering\let\newline\\\arraybackslash\hspace{0pt}}m{#1}}
\newcolumntype{R}[1]{>{\raggedleft\let\newline\\\arraybackslash\hspace{0pt}}m{#1}}
\begin{document}

\title{Provenance Threat Modeling}

\author{\IEEEauthorblockN{Oluwakemi Hambolu, \\ Lu Yu, Jon Oakley \\ and Richard R. Brooks }
\IEEEauthorblockA{Department of Electrical and\\Computer Engineering\\
Clemson University\\
Clemson, SC 29634\\
Email: ohambol@clemson.edu}
\and
\and
\IEEEauthorblockN{Ujan Mukhopadhyay \\ and Anthony Skjellum}
\IEEEauthorblockA{Department of Electrical and\\Computer Engineering\\
Auburn University\
Auburn, AL 36849\\
Email: uzm0002@tigermail.auburn.edu }}


%


\maketitle

\begin{abstract}
Provenance systems are used to capture history metadata, applications include ownership attribution and determining the quality of a particular data set. Provenance systems are also used for debugging, process improvement, understanding data proof of ownership, certification of validity, etc. The provenance of data includes information about the processes and source data that leads to the current representation. In this paper we study the security risks provenance systems might be exposed to and recommend security solutions to better protect the provenance information.
\end{abstract}


%
\IEEEpeerreviewmaketitle

\section{Introduction}
\noindent In many application areas like e-Science, detailed information about the origin of data is needed. This information is crucial in deciding if the data can be trusted, how it can be integrated into other sources, determining where an error may have been made, and deciding how to give credit to its originators when reusing it~\cite{Moreau:2011:OPM:1967762.1967931}. This kind of information is known as provenance. Provenance, a kind of metadata sometimes called ``lineage" or ``pedigree" has been described in various terms depending on the domain. In database system, the authors of~\cite{Buneman01whyand} defined it as a description of the origin of a piece of data and the process by which it arrived in a database. W3C PROV's specification~\cite{w3c-prov-primer} defines provenance as the information about entities, activities, and people involved in producing an artifact, which can be used to form assessments about its quality, reliability, or trustworthiness. In application to digital scientific data, provenance is an important component in broadening, sharing, and reusing scientific data~\cite{KarmaMisc}. Provenance is useful in validating results, failure tracing, and reproducibility~\cite{suriarachchi2015komadu}.

Provenance of a data product encompasses data acquisition, compilation methods, conversions, transformations, and analyses. There are two forms of navigating provenance: moving backward to discover ancestor products or transformations, or moving forward to discover descendant products or transformations~\cite{Bose2005LRS1057977.1057978}. 

Provenance information exists in many applications with a diverse variety of tools available to keep track of it. This paper concentrates on how to maintain provenance metadata securely. Most provenance tools used do not provide security guarantees that are needed to improve trust. We proposed ways to provide integrity, confidentiality, and privacy assurances in data provenance systems.

The remainder of this paper is organized as follows: Section~\ref{sec:back} gives background on provenance. The need for security in provenance systems is briefly discussed in Section~\ref{sec:security}. Finally, Section~\ref{sec:concl} presents our conclusion and future work.

\section{Background on Data Provenance}\label{sec:back}
\noindent Different approaches have been implemented to support data provenance in several domains.  Many domain areas like academic/research organizations and business establishments use provenance.  The provenance information is limited to the application domain, data representation model, or data processing facility~\cite{GD07}. 

In scientific domains, the authors in~\cite{SimmhanIucstr2005} and~\cite{Simmhan2005SDP1084805.1084812}, shows how sharing data and metadata across organizations has become the norm for strengthening the collaborative environment. Provenance has been used to provide data quality and attribution when using third-party data in the academic and research fields.  Publications are a common form of representing provenance for experimental data and results. They currently use uniform resource names like Handles, persistent URL (PURL), and Digital Object Identifiers (DOIs)~\cite{Arms2001URN374308.375358}, to cite the date used in experiments, enabling other researchers to relate the data's lineage to the actual data used. 

In the business domain, Simmhan et al.~~\cite{SimmhanIucstr2005} showed how large proportions of businesses deal with bad data especially when data are collected from different parts of the business and aggregated into the data warehouse. Data warehouse provides an integrated view of the history of the data from multiple sources. In this environment, provenance information can be used to trace the data in the warehouse back to the source from whence it was generated, as well as to trace the source of errors, enabling adequate corrections~\cite{Cui2003LTG775452.775456}.

Provenance information can be used in digital forensics which is the process of preserving, collecting, confirming, identifying, analyzing, recording, and presenting crime scene information~\cite{DBLPjournals/corr/abs-1211-4328}. This provides an audit trail to maintain the integrity of the information and give a strict chain of custody for the data. 

It is difficult to obtain provenance information in cloud forensics, a method of applying digital forensics to a cloud computing environment. Zawoad  and Hasan in ~\cite{DBLPjournals/corr/abs-1211-4328} discusses the challenges of providing provenance information in a cloud, because of the black-box nature of clouds and multi-tenant cloud models. They introduced the idea of building proofs of past data possession in the context of cloud storage to provide provenance for cloud forensics~\cite{DBLPjournals/corr/abs-1211-4328}. Zawoad et al. ~\cite{DBLPconf} proposed an open cloud forensics model that includes support for reliable digital forensics in the cloud.

According to~\cite{GD07}, the other application domains that may benefit from provenance are the interactive statistical environment, visualization, and Knowledge discovery in databases (KDD). Given the wide range of application domains that would benefit from provenance information, it is necessary to study the existing provenance tools available.

\section{Security Needs for Provenance Systems}\label{sec:security}
\noindent Significant research in provenance focuses on recording, managing, and using provenance information, but little work has been devoted to security goals such as confidentiality, integrity, availability and privacy~\cite{5992138}. 

Provenance metadata faces a number of security threats, including active attacks from intruders. An intruder's goal might be to compromise the provenance data, which in some cases might be more sensitive than the data itself~\cite{6297930}. For example, in digital forensics, provenance information is used in court, this information might be more sensitive than the data. Digital data can easily be copied, erased or tampered with, unlike physical documents. Even insiders might have some financial or strategic reasons to violate privacy and confidentiality by altering the history of the data~\cite{hasan2009protecting}. Provenance metadata becomes vulnerable to illegal alterations as it passes through untrusted environments.

This brings about the need to provide a model for securing the provenance; that is, ``provenance of provenance". 
It is challenging to make provenance records trustworthy because of the need to guarantee completeness~\cite{hasan2009protecting}.  Security in a provenance system can be viewed as maintaining the following services:
\begin{enumerate}
\item \textit{Integrity}: Provenance that cannot be forged or altered~\cite{hasan2009protecting}; provenance can only be modified by authorized parties~\cite{Stallings:1995:NIS:193189}.
\item \textit{Availability}: Allowing auditors to easily verify the integrity~\cite{hasan2009protecting}; provenance should be available to authorized users~\cite{Stallings:1995:NIS:193189}.
\item  \textit{Confidentiality}: Allowing only authorized parties to read the provenance~\cite{hasan2009protecting}.
\item \textit{Efficiency}: Provenance systems with low overheads~\cite{hasan2009protecting,Stallings:1995:NIS:193189}.
\item \textit{Authentication}: Provenance is correctly identified~\cite{Stallings:1995:NIS:193189}.
\item \textit{Nonrepudiation}: Neither sender not receiver can deny the existence of the provenance~\cite{Stallings:1995:NIS:193189}.
\item \textit{Access control}: Access to provenance is controlled and limited~\cite{Stallings:1995:NIS:193189}.
\end{enumerate}

\subsection{Attack Threat and Risk}
\begin{figure*}[!]
\begin{center}
\includegraphics[width=.6\textwidth]{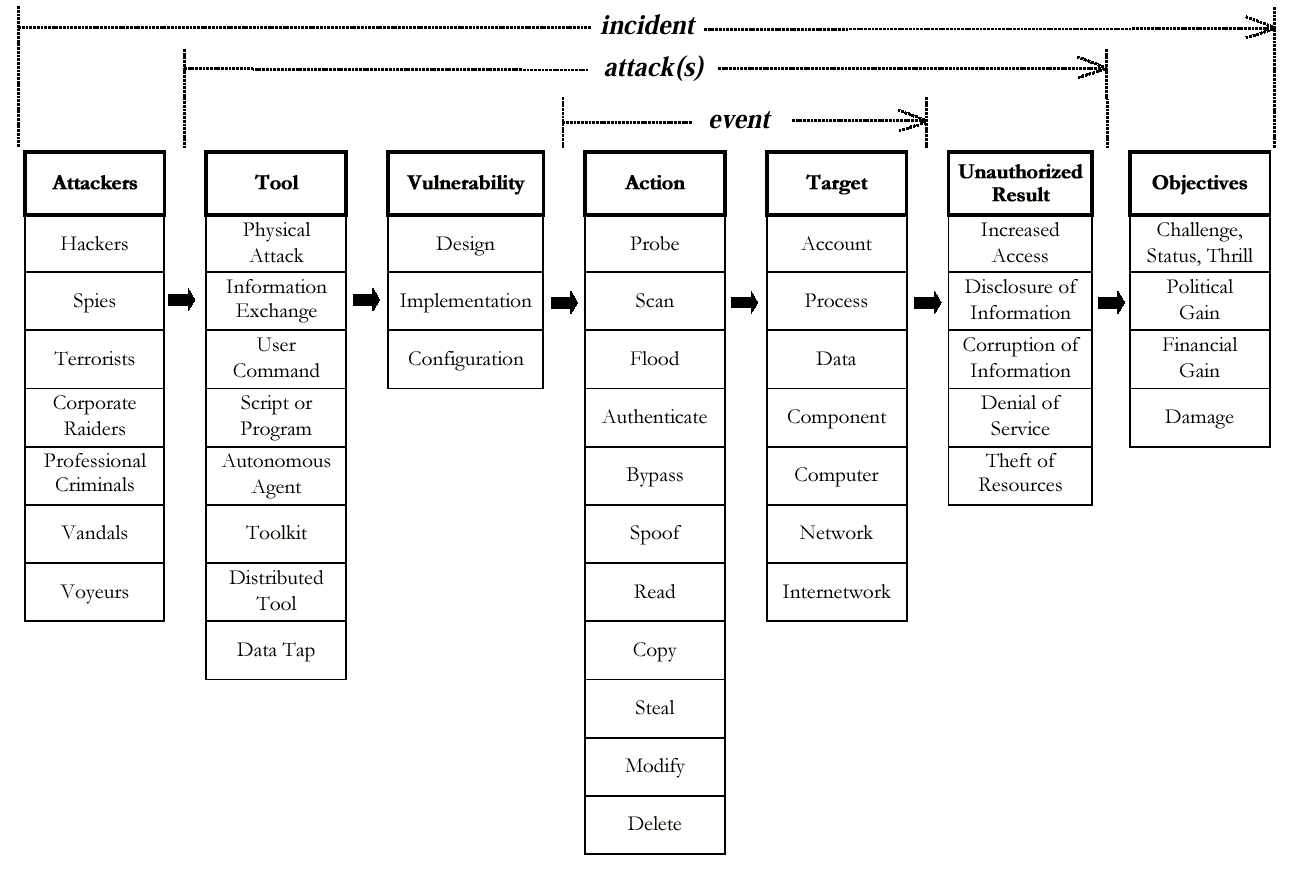}
\caption{Taxonomy of security incidents, adapted from~\cite{Howard98acommon}.}
\label{Fig:TaxScheme}
\end{center}
\end{figure*} 
\noindent The taxonomy presented in Figure~\ref{Fig:TaxScheme} classifies security incidents. Each incident, composed of one or more attacks, show how an attacker abuses the system to achieve an objective. The attacker uses one or more tools to exploit system vulnerabilities. Events show ways of exploiting the vulnerabilities~\cite{Brooks:2004:DST:1204608}.
\subsubsection{Threat Modeling}
\noindent Threat modeling analyzes system security from the intruder's perspective~\cite{caso2006application}. 
Exploiting the vulnerabilities of a system to create unauthorized results is known as an attack~\cite{Brooks:2004:DST:1204608}. Based on the security need of a system, the four general classes of attacks~\cite{Brooks:2004:DST:1204608,Stallings:1995:NIS:193189} can be applied to a provenance system :
\begin{enumerate}
\item \textit{Interruption}: Availability of provenance is disrupted.
\item \textit{Fabrication}: Insertion of bad/malicious provenance data.
\item \textit{Interception}: Unauthorized access to provenance.
\item \textit{Modification}: Unauthorized tampering of provenance.
\end{enumerate}
The threats to provenance systems considered can be classified under two forms: internal threats and external threats:
\begin{enumerate}
\item \textit{Internal Threats}:

Internal threats come from individuals with legitimate access to the system. These threats are hard to detect and mitigate because the attackers might know what to look for to avoid detection. Insiders can easily insert bad data and modify the existing data to suit their needs~\cite{DickyH}.
\item \textit{External Threats}:

External threats are usually performed by hackers/cracker, saboteurs and thieves~\cite{DickyH}. The most common way of gaining access to the system is password theft. ~\cite{DickyH} provides multiple ways an intruder can gain access to an authorized account, such as social engineering, exploiting system vulnerabilities, eavesdropping on the network traffic, etc. Once an intruder gains access to the system, they can attack/ misuse the information as they see fit.	
\end{enumerate}
\subsubsection{Risks}
\noindent According to~\cite{DickyH}, possible risks applicable to provenance systems include:
\begin{enumerate}
\item \textit{Unauthorized disclosure of information}: This includes disclosure of confidential, sensitive or embarrassing information, which leads to loss of credibility or reputation.
\item \textit{Disruption of services}: As a results of the threats, resources might be unavailable when needed, causing loss of productivity. 
\item \textit{Data tampering}: In the case of digital forensics, criminals or lawyers can tamper with the provenance information to suit their needs.
\end{enumerate}
\subsection{Security Solutions}
\noindent Suggested mitigation solutions against the threats are:
\begin{enumerate}
\item \textit{Cryptographic Digital Signatures}:  Cryptographic digital signature is a cryptographic value calculated from the data and a secret key only known by the signer. It binds the signer to the digital data~\cite{AnonyMisc}. According to~\cite{AnonyMisc}, digital signatures provide:
\begin{itemize}
\item \textit{Data authentication}: Using cryptographic signatures in provenance systems can ensure data is created by the signer and no one else.
\item \textit{Data integrity}: Cryptographic signatures make it possible to determine if the data has been accessed or modified by an attacker.
\item \textit{Non-repudiation}: When bad provenance data is found, the owner of the data can be verified/ identified.
\end{itemize}

\item \textit{Access Control}: Access control is the process of ensuring that authenticated users can access only what they are authorized to~\cite{ScottMisc}. Having appropriate access control set up will prevent unauthorized access to the provenance data. Access control can be divided into two parts:
\begin{itemize}
\item \textit{Authentication}: This is the process of identifying the identity of a user. The aim is to verify if the user attempting to gain access is allowed to do so~\cite{ScottMisc}. Typically, one or more of the following are used for access control: knowledge (such as password) , token (such as a key), or biometrics (such as fingerprint)~\cite{Brooks:2004:DST:1204608}.
\item \textit{Authorization}: This is the process of determining the access level of an authorized user~\cite{ScottMisc}.
\end{itemize}
Access control is insufficient to providing needed protection on the provenance data when considering internal threats~\cite{DickyH}. 
\item \textit{Incorporating cryptocurrency primitives}: A novel method to secure provenance metadata, will be to store provenance entries in a distributed ledger system, like the Bitcoin blockchain. The blockchain is a peer-to-peer distributed public ledger in which every transaction is registered~\cite{Nakamoto_bitcoina} making it easy to track the ownership. It consists of a distributed, chronological chain of blocks, with a linear path from the first block to the current block. The blockchain allows transactions, or other data, to be securely stored and verified without a centralized authority~\cite{Nakamoto_bitcoina}. One benefit of the blockchain is that data can be verified and time-stamped, creating an audit trail. 

The security of the blockchain is guaranteed by maintaining a cryptographically signed chain of secure hash values. Since this chain is stored at multiple sites and is also being continuously updated, it will be functionally impossible for this chain to be manipulated by fraudsters.
With this ability, the blockchain can ensure the integrity and security of provenance data. 

In order to subvert such a system, an intruder would have to gain control of all copies of the ledger and also be able to forge signatures from all parties providing the inputs. The technical details for this model are beyond the scope of this paper.
\end{enumerate}

\section{ Conclusion and Future Work}\label{sec:concl}
\noindent In this paper, we introduced data provenance, the explicit representation of the origin of data. We briefly discussed the background on data provenance in e-science. 

Using provenance as a basis for decision making largely depends upon the trustworthiness of provenance~\cite{SimmhanIucstr2005}, which can be increased if the system that store and represent provenance metadata provide security guarantees.

Given that provenance metadata face a number of security threats, from outsiders and insiders, security guarantees need to be provided by the systems that store the provenance information. This will avoid data corruption or manipulation, which will increase trust and data sharing, helping, for example, reviewers, funding agencies, and scientists ensure reproducibility of published scientific results. Leveraging  cryptocurrency primitives will ensure security, integrity, and confidentiality of the provenance information, which will be hard to subvert. As future work, we will present a detailed security model for securing provenance metadata using cryptocurrency primitives.


\ifCLASSOPTIONcompsoc
  \section*{Acknowledgments}
\else
  \section*{Acknowledgment}
\fi
\noindent This material is based upon work supported by the National Science Foundation under Grants Nos.~ 1547164 and 1547245.  Any opinions, findings, and conclusions or recommendations expressed in this material are those of the authors and do not necessarily reflect the views of the National Science Foundation.



\bibliographystyle{IEEEtran}
\bibliography{./bibtex}
%
%
%

\end{document}